\documentclass{aa}
\usepackage{graphicx}

\begin{document}

\title{One-sided jet at milliarcsecond scales in \object{LS~I~+61$^{\circ}$303}}

\author{M. Massi\inst{1}
\and M. Rib\'o\inst{2}
\and J.M. Paredes\inst{2}
\and M. Peracaula\inst{2}
\and R. Estalella\inst{2}
}

\offprints{M. Massi, \\ \email{mmassi@mpifr-bonn.mpg.de}}

\institute{Max Planck Institut f\"ur Radioastronomie, Auf dem H\"ugel 69, 
D-53121 Bonn, Germany\\
\and Departament d'Astronomia i Meteorologia, Universitat de Barcelona, Av. Diagonal 647, E-08028 Barcelona, Spain
}

\date{Received / Accepted}

\abstract{
We present Very Long Baseline Interferometry (VLBI) observations of the high
mass X-ray binary \object{LS~I~+61$^{\circ}$303}, carried out with the European
VLBI Network (EVN). Over the 11 hour observing run, performed $\sim$10
days after a radio outburst, the radio source showed a constant flux density,
which allowed sensitive imaging of the emission distribution. The structure in
the map shows a clear extension to the southeast. Comparing our data with
previous VLBI observations we interpret the extension as a collimated radio jet
as found in several other X-ray binaries. Assuming that the structure is the
result of an expansion that started at the onset of the outburst, we derive an
apparent expansion velocity of $0.003$ c, which, in the context of Doppler
boosting, corresponds to an intrinsic velocity of at least $0.4$ c for an
ejection close to the line of sight. From the apparent velocity in all
available epochs we are able to establish variations in the ejection angle
which imply a precessing accretion disk. Finally we point out that
\object{LS~I~+61$^{\circ}$303}, like \object{SS~433} and \object{Cygnus~X-1},
shows evidence for an emission region almost orthogonal to the relativistic
jet.
}

\maketitle

\section{Introduction} \label{introduction}

\object{LS~I~+61$^{\circ}$303} is an X-ray binary system associated with the
galactic plane variable radio source GT~0236+610 discovered by Gregory \&
Taylor (\cite{gregory78}). Optical observations (Hutchings \& Crampton
\cite{hutchings81}) show that the system is composed of a neutron star and an
early type, rapidly rotating B0V star with a stable equatorial disk and mass
loss. Spectral line observations of the radio source give a distance of
2.0$\pm$0.2~kpc (Frail \& Hjellming \cite{frail91}).

One of the most unusual aspects of its radio emission is the fact that it
exhibits two periodicities: a 26.5 day periodic nonthermal outburst (Taylor \&
Gregory \cite{taylor82}, \cite{taylor84}) and a 1584 day ($\sim$ 4 years)
modulation of the outburst peak flux (Gregory et~al. \cite{gregory99}). The
26.5 day periodicity corresponds to the orbital period of the binary system
(Hutchings \& Crampton \cite{hutchings81}). This periodicity has also been
detected in UBVRI photometric observations (Mendelson \& Mazeh
\cite{mendelson89}), in the infrared domain (Paredes et~al. \cite{paredes94}),
in soft X-rays (Paredes et~al. \cite{paredes97}) and in the H$\alpha$ emission
line (Zamanov et~al. \cite{zamanov99}). The 4 year modulation has been observed
as well in the H$\alpha$ emission line (Zamanov et~al. \cite{zamanov99}).

Simultaneous X-ray and radio observations show that the X-ray outbursts occur
at the periastron passage while, on the contrary, the strongest radio outbursts
are always delayed with respect to that (Taylor et~al. \cite{taylor96}; Gregory
et~al. \cite{gregory99}). Both, the presence of two periodicities (at 26.5 days
and $\sim 4$ years) and the delay between radio and X-ray outbursts, are well
explained in the framework of an accretion scenario of a precessing neutron
star in a highly ($e>0.4$) eccentric orbit (Gregory et~al. \cite{gregory89};
Gregory et~al. \cite{gregory99}). The accretion rate in an eccentric orbit
within the equatorial wind of the Be star has two peaks: the highest peak
corresponds to the periastron passage and the second, lower amplitude, peak
occurs when the relative velocity of the neutron star and the Be star wind is
at a minimum. For supercritical accretion, matter is ejected outwards in two
jets perpendicularly to the accretion disk plane. Near periastron, inverse
Compton losses are severe (due to the proximity to the Be star): X-ray
outbursts are expected but not radio ones. For the second accretion peak, the
neutron star is much farther from the Be star and both inverse Compton losses
and wind opacity are lower, the electrons can propagate out of the orbital
plane and we observe the radio outburst. The precession of the disk gives rise
to the $\sim$ 4 year modulation (Gregory et~al. \cite{gregory89}; Taylor et~al.
\cite{taylor92}; Massi et~al. \cite{massi93}; Mart\'{\i} \& Paredes
\cite{marti95}; Gregory et~al. \cite{gregory99}). The presence of an accretion
disk is also invoked by Mendelson \& Mazeh (\cite{mendelson89}) to explain
details of the optical light curve. Liu et~al. (\cite{liu00}) explain the
variation of the H$\alpha$ emission with orbital phase as varying irradiation
of the Be star's circumstellar disk by the X-ray emission from the neutron
star's accretion disk.

\begin{figure}
\resizebox{\hsize}{!}{\includegraphics{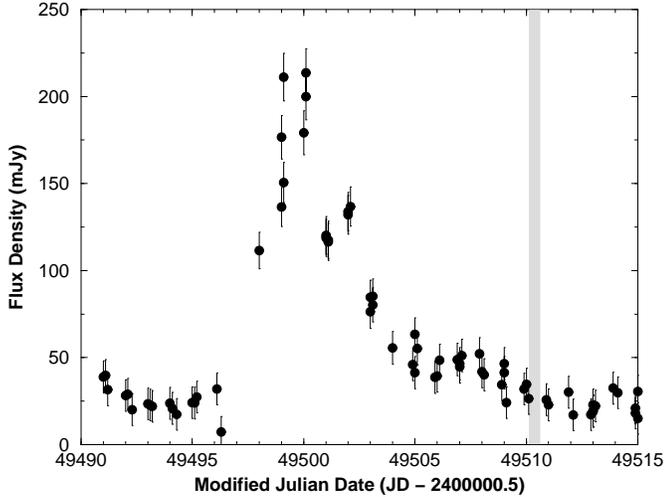}}
\caption{Radio light curve of \object{LS~I~+61$^{\circ}$303} obtained with the GBI at 8.4~GHz (Strickman et~al. \cite{strickman98}). The shaded area indicates the time interval during which our VLBI data were taken}
\label{gbi}
\end{figure}

However, the luminosity of \object{LS~I~+61$^{\circ}$303} in the X-ray range
(1--40~keV), is only $L_{\rm X}\simeq10^{35}$~erg~s$^{-1}$ (Maraschi \& Treves
\cite{maraschi81}), which is three orders of magnitude lower than the Eddington
limit. Moreover, \object{LS~I~+61$^{\circ}$303} is the most promising candidate
for the optical counterpart of the $\gamma$-ray source 3EG~J0241+6103 (Hartman
et~al. \cite{hartman99}), with a luminosity $L_{\rm
\gamma}\simeq10^{37}$erg~s$^{-1}$. The fact that \object{LS~I~+61$^{\circ}$303}
has the bulk of its energy shifted from X-ray to $\gamma$-ray wavelengths is
not understood in the context of the supercritical accretion model, showing the
necessity for alternative models. These models suggest that the 26.5~d outburst
events are produced by energetic electrons accelerated in the shock boundary
between the relativistic wind of a young non-accreting pulsar and the wind of
the Be star. In this case the 4 year modulation would be explained due to
cyclic variations of the Be star envelope (Maraschi \& Treves
\cite{maraschi81}; Tavani \cite{tavani94}; Tavani et~al. \cite{tavani98};
Goldoni \& Mereghetti \cite{goldoni95}; Zamanov \cite{zamanov95}; Taylor et~al.
\cite{taylor96}).

The recent discovery of the microquasar LS~5039 (Paredes et~al.
\cite{paredes00}) brings new credibility to the accretion model. LS~5039 is
also subluminous in the X-ray range (even more than
\object{LS~I~+61$^{\circ}$303}) and also shows the same puzzling behavior,
having $L_{\gamma} > L_{\rm X}$. Therefore, LS~5039 and
\object{LS~I~+61$^{\circ}$303} could be the first two examples of a new class
of X-ray binaries with powerful $\gamma$-ray emission. In the case of LS~5039,
an ejection process fed by an accretion disk is clearly proved by a map
obtained with the Very Long Baseline Array (VLBA), which shows bipolar jets
emerging from a central core. Its high $L_{\gamma}$ is tentatively explained by
inverse Compton scattering. In \object{LS~I~+61$^{\circ}$303}, although several
VLBI observations show a complex source extending over a few
milliarcseconds (Massi et~al. \cite{massi93}; Peracaula et~al.
\cite{peracaula98}; Paredes et~al. \cite{paredes98}; Taylor et~al.
\cite{taylor00}), such a clear jet structure has never been observed.

In this paper we report on EVN observations probing structures on scales of
tens of milliarcseconds, larger than probed by previous observations. We
clearly detect on this scale an elongation of the emission
(Sect.~\ref{observations}), which we interpret in Sect.~\ref{discussion}, as a
one-sided jet. We determine its apparent expansion velocity and derive the
intrinsic velocity considering Doppler boosting. Our conclusions are given in
Sect.~\ref{conclusions}.

\begin{figure}
\resizebox{\hsize}{!}{\includegraphics{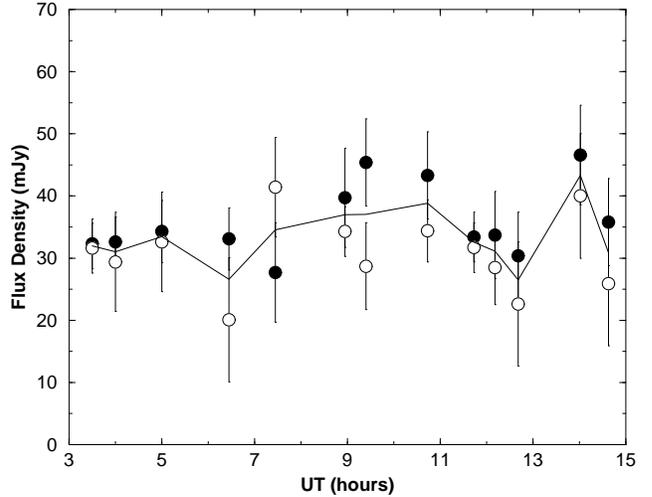}}
\caption{Effelsberg flux density measurements of \object{LS~I~+61$^{\circ}$303} at 5~GHz obtained during our VLBI observations. LCP and RCP flux density measurements are represented by filled and open circles respectively. The solid line corresponds to the total flux density}
\label{eff}
\end{figure}

\section{Observations and results} \label{observations}

Our observations of \object{LS~I~+61$^{\circ}$303} were carried out during 11
hours on 1994 June 7, at 5~GHz, using four antennas of the EVN, namely the
Effelsberg 100m, Medicina 32m, Noto 32m and Onsala 25m telescopes. The data
were recorded, in left hand circular polarization, using the Mark III mode A
recording system, corresponding to a total bandwidth of 56~MHz. The correlation
of the data was done on the Mark III correlator operated by the
Max-Planck-Institut f\"ur Radioastronomie in Bonn.

Our EVN observations were made within a time interval during which a
multiwavelength monitoring campaign was conducted that included simultaneous
radio, optical, infrared, and hard X-ray/$\gamma$-ray observations and covered
almost three orbital cycles of \object{LS~I~+61$^{\circ}$303} (Strickman et~al.
\cite{strickman98}). As can be seen from the radio light curve taken with the
Green Bank Interferometer\footnote{The Green Bank Interferometer is a facility
of the USA National Science Foundation operated by the NRAO in support of NASA
High Energy Astrophysics programs.} (GBI) as part of that campaign
(Fig.~\ref{gbi}), our observations took place about 10 days after the radio
outburst maximum (i.e. almost 13 days after the onset of the outburst). Total
flux density observations of \object{LS~I~+61$^{\circ}$303}, carried out by the
100m telescope of Effelsberg at the beginning of each VLBI scan, are shown in
Fig.~\ref{eff}. The variations of left and right hand circular polarization
flux are within their error bars. This lack of significant circular
polarization agrees with previous results of Peracaula et al.
(\cite{peracaula97}). The total flux density is nearly constant and with an
average value of $34\pm5$~mJy.

\begin{figure}
\resizebox{\hsize}{!}{\includegraphics{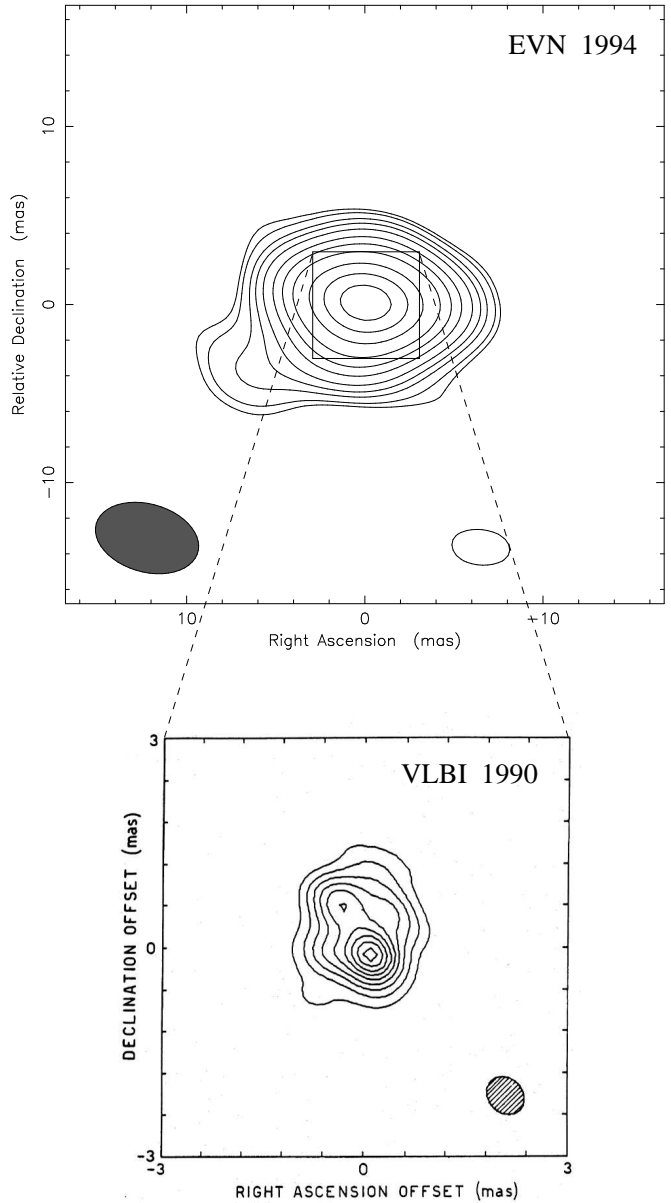}}
\caption{{\bf Top:} EVN uniform weighted map of \object{LS~I~+61$^{\circ}$303} at 5~GHz frequency obtained on 1994 June 7 (this paper). The contours are at $-$3, 3, 4, 6, 8, 11, 16, 22, 30, 45, 60 and 75 times the r.m.s. noise of 0.28~mJy~beam$^{-1}$. The filled ellipse in the bottom-left corner represents the FWHM of the synthesized beam, which is 5.9~mas$\times$3.8~mas at a P.A. of 74.2$^{\circ}$. {\bf Bottom:} VLBI map (at 5~GHz) by Massi et~al. (\cite{massi93}), obtained on 1990 June 6, with a resolution of 0.6~mas $\times$ 0.5~mas}
\label{maps}
\end{figure}

In our EVN observations, \object{LS~I~+61$^{\circ}$303} was clearly detected on
all baselines. The data were reduced using standard procedures within the AIPS
and Difmap software packages. The EVN uniform weighted map is shown at the top
of Fig.~\ref{maps}. It exhibits an extended structure elongated to the
southeast. The elongation is not artificially created by the beam, in fact the
elongation is at P.A.$\simeq120^{\circ}$ while the beam, of size
5.9~mas~$\times~$3.8~mas, has P.A.=74.2$^{\circ}$. The flux density recovered
in the cleaning process amounts to $\sim35$~mJy.

In Table~\ref{models} we show the parameters obtained after fitting several
models to the $(u,v)$ data. The best fit using a single component is obtained
with an elliptical Gaussian (Model~1 in Table~\ref{models}). A circular
Gaussian or a point source give higher values of $\chi^2_r$. The elliptical
Gaussian has a P.A. consistent with the observable elongation in the map.
However, there is still residual emission not accounted by Model~1, in the
direction of the elongation.

\begin{table*}
\begin{flushleft}
\caption[]{Parameters of different models fitted to the data}
\label{models} 
\begin{tabular}{lllllllll}
\hline\hline \noalign{\smallskip}
Model & $\chi^2_r$ & Components & Flux Density & r     & $\theta$  & Maj. axis & Min. axis & P.A. \\
      &            &            & (mJy)        & (mas) & ($\degr$) & (mas)     & (mas)     & ($\degr$) \\
\noalign{\smallskip} \hline\hline \noalign{\smallskip}
1     & 0.798 & Elliptical Gaussian & 33.0 & 0.04 & 3   & 3.7 & 2.5 & 123 \\
\noalign{\smallskip} \hline \noalign{\smallskip}
2     & 0.786 & Elliptical Gaussian & 21.6 & 0.5  & 135 & 5.5 & 3.6 & 123 \\
      &       & Point source        & 13.1 & 0.5  & -43 &     &     &     \\
\noalign{\smallskip} \hline\hline
\end{tabular}
\end{flushleft}
\end{table*}

In previous VLBI observations (Paredes et~al. \cite{paredes98}) the best fits
to the $(u,v)$ data were consistent with an unresolved core plus a Gaussian
halo. Hence, we have added a point source to Model~1, and we have obtained a
slightly improved fit to the data (Model~2 in Table~\ref{models}). It is
important to note that, in this model, the elliptical Gaussian conserves its
P.A. and moves towards the direction of the elongation, while the point source
moves in the opposite direction. We have tried several different initial
conditions and always the fit converged to the same solution. Finally, for a
distance of 2.0~kpc (Frail \& Hjellming \cite{frail91}), the brightness
temperature of the components of the models listed in Table~\ref{models} are in
the range of $T_{\rm B}=10^{7}$-$10^{8}$~K, which are characteristic of
non-thermal emission.

\begin{table*}
\begin{flushleft}
\caption[]{Summary of published VLBI observations of \object{LS~I~+61$^{\circ}$303} at 5~GHz. Phase is computed according to $P=26.4917$~d and $T_0={\rm JD}~2\,443\,366.775$}
\label{history}
\begin{tabular}{llllllll}
\hline\hline \noalign{\smallskip}
Epoch       & Array       & Phase       & State                   & Flux Density       & Size        & P.A.                 & Expansion \\
            &             &             &                         & (mJy)              & (mas)       & ($^{\circ}$)         & Velocity/$c$ \\
\noalign{\smallskip} \hline\hline \noalign{\smallskip}
1987 Sep 25$^a$ &         & 0.59        & quiescent               & \phantom{1}40--54  & $3.2\pm0.9$ &                      & \\ 
            & EVN         &             &                         &                    &             &                      & $\leq0.002$ \\
1987 Oct 1$^a$  &         & 0.81        & burst, decaying         & 260--200           & $1.6\pm1.2$ &                      & \\ 
\noalign{\smallskip} \hline \noalign{\smallskip}
1990 Jun 6$^b$  & EVN + VLA  & 0.74     & burst, decaying         & 244--205           & $\sim2$     & $\sim135$, $\sim$ 30 & $\sim0.002^1$ \\
\noalign{\smallskip} \hline \noalign{\smallskip}
1992 Jun 8$^c$  & Global VLBI & 0.42    & quiescent, minioutburst & \phantom{1}35      & 0.5--2      & $\sim160$            & 0.06 \\ 
\noalign{\smallskip} \hline \noalign{\smallskip}
1993 Sep 9$^d$  &         & 0.69        & burst, variable         & \phantom{1}76--131 & 2--3        &                      & \\ 
            & Global VLBI &             &                         &                    &             &                      & 0.007 \\
1993 Sep 13$^d$ &         & 0.84        & quiescent               & \phantom{1}60      & $\sim7$     & $\sim120$            & \\ 
\noalign{\smallskip} \hline \noalign{\smallskip} 
1994 Jun 7$^e$  & EVN     & 0.92        & quiescent               & \phantom{1}34      & $\sim6$     & $\sim120$            & 0.003 \\ 
\noalign{\smallskip} \hline \noalign{\smallskip}
1999 Sep 16-17$^f$ & HALCA + & 0.69     & burst, variable         & 140                & $\sim4$     &                   & $\sim0.002^2$ \\ 
            & Global VLBI &             &                         &                    &             &                      & \\
\noalign{\smallskip} \hline\hline \noalign{\smallskip}
\end{tabular}
$^a$~Taylor et~al. \cite{taylor92}; $^b$~Massi et~al. \cite{massi93}; 
$^c$~Peracaula et~al. \cite{peracaula98}; $^d$~Paredes et~al. 
\cite{paredes98}; $^e$~This paper; $^f$~Taylor et~al. \cite{taylor00} \\
$^1$~Recalculated according to new outburst peak ephemerides (Gregory
et~al. \cite{gregory99}), for the extended structure at P.A.=135$^{\circ}$ \\
$^2$~There is not enough resolution at P.A. $\sim140$ to know if there was expansion in that direction
\end{flushleft}
\end{table*}

\section{Discussion} \label{discussion}

\subsection{Morphology} \label{morphology}

In view of the models that can fit the data, the source in Fig.~\ref{maps} can
be explained as a central core (point source of Model~2 in Table~\ref{models})
and a one-sided jet (elliptical Gaussian in the same model). If we assume that
the Gaussian component is the result of an expanding source that originated at
the onset of the major outburst, that is about 13 days before our observations,
and that the expansion velocity of the jet is constant, the derived velocity,
taking into account the size of the Gaussian and the offset to the point
source, is $\sim0.3$~mas~d$^{-1}$. This corresponds, at a 2.0~kpc distance, to
a projected expansion velocity of $\sim0.003$ c. 

The most interesting aspect of the EVN map is that, for the first time, we
detected asymmetric emission in the southeast direction. The morphology is
quite similar to that observed by Stirling et~al. (\cite{stirling00}, Fig.~4)
in \object{Cygnus~X-1} at 15~GHz: a central source with a small but clear
elongation. Is this morphological analogy enough to interpret the structure in
the EVN map as a one-sided jet? In the case of \object{Cygnus~X-1} the
interpretation of the small jet-like elongation has proved to be correct: a
later observation of this source at 8~GHz shows the elongation developed in an
extended jet.

\object{LS~I~+61$^{\circ}$303} has been observed several times with mas
resolution. In Table~\ref{history} we report previous VLBI observations
focusing on two special items: the position angle (P.A.) of any extended
feature present and its expansion velocity. As can be seen, an extended feature
with P.A. from 120$^{\circ}$ to 160$^{\circ}$ seems to be present in other
maps. While the identification is perhaps ambiguous in Paredes et al.
\cite{paredes98} because the extension is coincident with the direction of the
major axis of the beam, the observations by Peracaula et al. (1998), report on
one extended feature with comparable P.A. with that of the feature in our map.
The reason why in the past this extended structure was never clearly associated
to a jet is because an internal structure at another P.A. ($\simeq$
30$^{\circ}$) makes the morphology confusing. To better understand this point
we show at the bottom of Fig.~\ref{maps} the map by Massi et~al.
(\cite{massi93}). The VLBI map resolves two components (at
P.A.$\simeq$30$^{\circ}$) inside the extended structure (at P.A.$\simeq$
135$^{\circ}$). The extended structure is asymmetric as well as that in our
map, but towards the northwest direction. This is probably a projection effect
corresponding to a different ejection angle due to precession
(Sect.~\ref{precession}). In Paredes et~al. (\cite{paredes98}) the double
internal source is not resolved, but the authors propose the presence of an
unresolved component stable and not participating in the flaring process,
inside an extended structure at P.A.$\simeq$ 120$^{\circ}$. These two
components are again resolved in the recent VLBI observation using HALCA
(Taylor et~al. \cite{taylor00}). While in the past this double internal source
was supposed to be the jet, in light of the EVN map we propose that:

\begin{enumerate}

\item The double source at P.A. of $\simeq$ 30$^{\circ}$ is not the jet.
Instead, it is almost orthogonal to the jet. It has a size of 2~AU and
therefore is comparable with the orbital size (1.4~AU).

\item The jet is identified with the structure with P.A. in the range 
120$^{\circ}$--160$^{\circ}$ seen in at least three VLBI observations, and 
best defined in our EVN map where it has a size of $\sim8$~AU.

\end{enumerate}

An emitting region perpendicular to the jet has been observed indeed also in
\object{SS~433} by Paragi et~al. (\cite{paragi99}) and perhaps in 
\object{Cygnus~X-1} by Stirling et~al. (\cite{stirling00}) and preliminary
interpreted as shocked gas in the orbital/accretion plane. If this
interpretation holds true also for \object{LS~I~+61$^{\circ}$303} the strong
wind opacity close to the periastron, discussed in the introduction, certainly
severely affects the morphology of such emission. For the jet, however, wind
opacity effects are negligible. The jet is generated by outbursts occurring at
quite a displaced orbital phase with respect to the periastron. Moreover, the
size of the jet is much greater than the orbital size.

\subsection{One-sided jet and Doppler boosting} \label{doppler}

It is well known that while some extragalactic radio sources have two jets, for
some others only one jet is observed. The unification model for AGN (see a
review in Urry \& Padovani \cite{urry95}) assumes that all of them represent
the same class of objects (all having two jets) and their different appearance
depends on different observing angles. Let us assume a symmetric ejection of
two jets at velocity $\beta$ (i.e. expressed as fraction of $c$). The two jets,
approaching and receding, move at an apparent velocity $\beta_{\rm a,r}$
related to the intrinsic $\beta$ by (Rees \cite{rees66}; Mirabel \&
Rodr\'{\i}guez \cite{mirabel94})
\begin{equation}
\beta_{\rm a,r}={\beta \sin\theta\over 1\mp \beta\cos\theta},
\label{e1}
\end{equation}
where $\theta$ is the angle between the direction of motion of the ejecta and
the line of sight. 

Following the method of Mirabel \& Rodr\'{\i}guez (\cite{mirabel94}) one can
determine the quantity $\beta \cos \theta$ by means of the ratio of flux
densities from the approaching and receding jets,
\begin{equation}
{S_{\rm a}\over{S_{\rm 
r}}}=\left({1+\beta\cos\theta\over1-\beta\cos\theta} \right)^{k-\alpha}.
\label{e4}
\end{equation}
where $\alpha$ is the spectral index of the emission ($S\propto \nu^{\alpha}$)
and $k$ is 2 for a continuous jet and 3 for discrete condensations.

In our case we deal with one jet only. However, we can determine the lower
limit
\begin{equation}
\beta\cos\theta> {\big({S_{\rm a}^{\rm
peak}/3\sigma}\big)^{1/(k-\alpha)}-1 \over \big({S_{\rm a}^{\rm
peak}/3\sigma}\big)^{1/(k-\alpha)}+1},
\label{e5}
\end{equation}
using the noise level ($\sigma$) of the map and the peak value of the
approaching component ($S_{\rm a}^{\rm peak}$). The spectral index a few days
after the outburst is $\alpha=-0.5$ (Strickman et~al. \cite{strickman98}). To
be consistent with the lowest limit we select \mbox{$k=3$}.

For our EVN map, with $\sigma=0.28$~mJy~beam$^{-1}$ and $S_{\rm a}^{\rm
peak}=21.6$~mJy~beam$^{-1}$, we have $\beta\cos\theta>0.43$.

Using Eq.~\ref{e1} we obtain
\begin{equation}
\beta=\sqrt{{(\beta\cos\theta)}^2+{\beta_{\rm a}}^2{(1-\beta\cos\theta)}^2},
\label{a6}
\end{equation}
which for $\beta_{\rm a}=0.003$ (Sect.~\ref{morphology}) and
$\beta\cos\theta>0.43$ gives an intrinsic velocity of $\beta>0.43$ and an
ejection angle $\theta\simeq0\degr$. If we consider a typical size of
$\sim10^{-2}$~pc for the jets in other known microquasars, the $\sim8$~AU size
of the \object{LS~I~+61$^{\circ}$303} jet would result in an angle of
$\theta\simeq0\fdg2$, compatible with our estimation.

As reviewed by Mirabel \& Rodr\'{\i}guez (\cite{mirabel99}), the expansion
velocities for microquasars range from $\sim0.1$ c to $\sim0.9$c
(\object{SS~433}, \object{\object{Cygnus~X-3}}, \object{GRS~1915+105},
\object{GRO~J1655$-$40}). The above determined lower limit of $0.4$c for
\object{LS~I~+61$^{\circ}$303} is therefore well within that range. Finally, we
note that the combination of the values estimated for $\theta$ and $\beta$
gives a Lorentz factor $\gamma=1.1$, and a Doppler factor $\delta_{\rm
a,r}={1\over \gamma (1 \mp \beta\cos\theta)}$, $\delta_{\rm a}=1.59$ for the
approaching jet and $\delta_{\rm r}=0.63$ for the receding one.

\subsection{Precession of the accretion disk} \label{precession}

The measured expansion velocities shown in Table~\ref{history} span a range of
0.002--0.007~$c$ and reach a value of $0.06~c$ at epoch 1992 June 8. On the
basis of our discussion in Sect.~\ref{morphology} and \ref{doppler}, we
interpret the observed expansion velocities as apparent transverse velocities,
$\beta_{\rm a}$, defined by Eq.~\ref{e1}. A possible explanation for the large
range of observed velocities is a variable intrinsic velocity $\beta$ and a
constant $\theta$. However, this is not supported by the observations available
up to now; moreover, for example, \object{SS~433} has shown a quite constant
velocity of $\beta=0.26$ for years. The alternative explanation is precession
of the jet, with the angle $\theta$ between the direction of the jet and the
line of sight being a function of time. Evidence for precession has been found
at least for \object{SS~433} and \object{Cygnus~X-1} (Brocksopp et~al.
\cite{brocksopp99}). Moreover, for \object{LS~I~+61$^{\circ}$303}, precession
of the jet has already been suggested to explain the 4~yr modulation of the
peak of the radio outbursts (Gregory et~al. \cite{gregory89}).

\begin{figure}
\resizebox{\hsize}{!}{\includegraphics{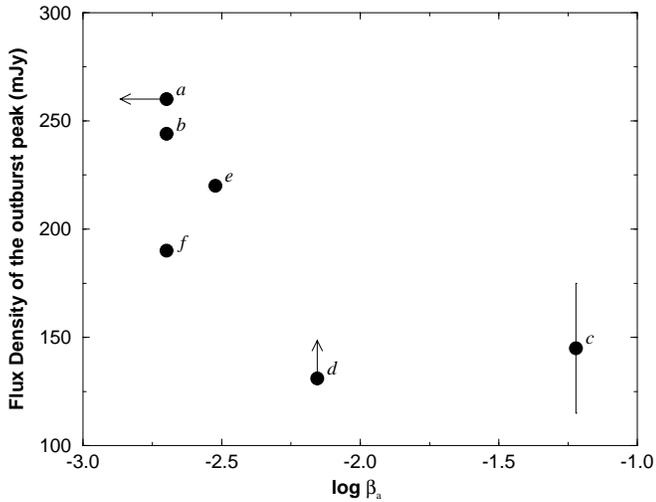}}
\caption{Flux density of the radio outburst peak versus $\log\beta_{\rm a}$ (see text). Each data point in this diagram is the result of the corresponding VLBI observation identified with letters $a$ to $f$ (see Table~\ref{history}). For observations $a$, $b$, $e$ and $f$ the flux density of the radio outburst peak was known from the same VLBI observations or from the GBI monitoring program. Point $d$ is lower limit of the outburst peak, while $c$ is taken from the average outburst peak flux modulation of Gregory et~al. (\cite{gregory99})}
\label{anticorr}
\end{figure}

If the latter assumption is correct, we would expect an anticorrelation between
the flux density of the radio outburst peak, and $\theta$ or $\beta_{\rm a}$.
In other words, when the jet is pointing directly towards us ($\theta$ small),
$\beta_{\rm a}$ is also small, and the flux density of the outburst peak is the
highest possible due to the Doppler boosting effect. On the contrary, when the
jet is not pointing directly towards us, $\theta$ and $\beta_{\rm a}$ increase,
and the flux density decreases. In Fig~\ref{anticorr} we show the flux density
of the radio outburst peak versus $\log\beta_{\rm a}$, for the VLBI
observations listed in Table~\ref{history}, taking $\beta_{\rm a}$ as the
observed expansion velocity. We can see from Fig~\ref{anticorr} that the
available data are consistent with the anticorrelation of the flux density of
the radio outburst peak and $\beta_{\rm a}$ predicted by our model, and thus
with the precession of the jet of \object{LS~I~+61$^{\circ}$303}.

Assuming that the the intrinsic velocity of the jet remains constant
($\beta\simeq0.4$) for all the observations listed in Table~\ref{history}, we
can derive that the maximum value of the ejection angle $\theta$, corresponding
to the maximum value of $\beta_{\rm a}=0.06$, is $\theta\simeq4\fdg5$. This
maximum angle is consistent with a moderate precession of the jet, and could be
the result of the precession of the accretion disk.

The range of values obtained for $\theta$ imply that the orbital plane of the
system would be close to the plane of the sky, i.e., the inclination of the
orbit, $i$, should be small. The other available information about $i$ comes
from spectroscopic optical and UV observations carried out by Hutchings \&
Crampton (\cite{hutchings81}). Although their observations reveal shell
absorption, this fact alone does not give information on the angle of
inclination, because the shell can cover the whole star and not be just an
equatorial bulge (Kogure \cite{kogure69}; Geuverink \cite{geuverink70}). We
note that the observations by Waters et~al. (\cite{waters88}) establish that,
together with a dense and slow disk-like wind around the equator, there exists
a high velocity, low density wind at higher latitudes up to the polar regions.
On the other hand, Hutchings \& Crampton (\cite{hutchings81}) obtain
$v\sin~i=360\pm25$~km~s$^{-1}$, where $v$ is the equatorial rotational velocity
of the Be star. As the maximum rotational velocity of a Be star (Hutchings
et~al. \cite{hutchings79}) is $v=630$~km~s$^{-1}$, this would result in a value
of $i\sim 35^{\circ}$. However, Hutchings \& Crampton (\cite{hutchings81})
comment that the velocity data are "extensive and unwieldy" and indeed the
value of $v\sin~i$ given above seems large if compared with the statistical
study of Be stars by Slettebak (\cite{slettebak82}). New observations would be
very useful to clarify this issue.

\section{Conclusions} \label{conclusions}

Our EVN observations at 5~GHz revealed that \object{LS~I~+61$^{\circ}$303} has
a jet-like elongation to the southeast on scales of tens of milliarcseconds.
This is interpreted as an approaching, Doppler boosted, jet of a symmetric
pair. The angle between the jet and the line of sight at the epoch of our
observation was close to zero. However, data at other epochs suggest the
existence of precession of the jet, and the accretion disk, of at least
$5^{\circ}$. This precession would change the angle of the ejecta and
consequently would vary both the apparent expansion velocity and the intensity
of the peak of the radio outbursts. New VLBI observations at several epochs are
necessary to correlate morphology, position angle, and expansion velocity, with
the 4~yr modulation of the peak of the radio outbursts (Gregory et~al.
\cite{gregory99}) and of the H$\alpha$ emission line (Zamanov et~al.
\cite{zamanov99}).

We derived a lower limit of $0.4~c$ for the intrinsic velocity of the radio
jet. This value is well within the range $0.1~c$ to $0.9~c$ found for
microquasars like \object{SS~433}, \object{Cygnus~X-3}, \object{GRS~1915+105}
and \object{GRO~J1655$-$40} (Mirabel \& Rodr\'{\i}guez \cite{mirabel99}).

Finally, in some microquasars emission almost orthogonal to the jet, i.e. along
the orbital plane, exists. This emission has clearly been observed in
\object{SS~433} (Paragi et~al. \cite{paragi99}) and it is barely visible in
\object{Cygnus~X-1} (Stirling et~al. \cite{stirling00}). As we showed in this
paper, \object{LS~I~+61$^{\circ}$303} could be a third case of this  kind of
sources.

\begin{acknowledgements}

We thank Karl Menten, Andrew Lobanov, Alok Patnaik, Eduardo Ros and
Giovanna Pugliese for useful discussions and comments.
We acknowledge detailed and very useful comments from L. Lara, the referee of this paper.
During this work, M.R. has been supported by two fellowships from CIRIT 
(Generalitat de Catalunya, ref. 1998~BEAI~200293 and 1999~FI~00199).
J.M.P. and M.P. acknowledge support from DGICYT grant PB97-0903 (Spain).

\end{acknowledgements}

\end{document}